\documentclass[10pt,a4paper,twoside]{article}
\usepackage[a4paper]{geometry}
\usepackage[T1]{fontenc}
\usepackage{mathpazo}
\usepackage[latin1]{inputenc}							
\usepackage{amsmath}									
\usepackage{xcolor,graphicx}
\definecolor{bl}{rgb}{0.0,0.2,0.6}
\definecolor{nicered}{rgb}{.647,.129,.149}

\usepackage{hyperref}
\hypersetup{
    bookmarks=true,         
    pdftoolbar=true,        
    pdfmenubar=true,        
    pdffitwindow=false,     
    pdfstartview={FitH},    
    pdftitle={A statistical physics perspective on criticality in financial markets},    
    pdfauthor={Thomas Bury},     
    pdfnewwindow=true,      
    colorlinks=true,       
    linkcolor=bl,          
    citecolor=blue ,        
    filecolor=red!90,      
    urlcolor=blue ,          
}

\usepackage{sectsty}
\usepackage[compact]{titlesec}
\titleformat{\section}{\color{nicered}\large\bf}{\thesection}{1em}{}
\titleformat{\subsection}{\color{nicered}\normalsize\bf}{\thesubsection}{1em}{}
\titleformat{\subsubsection}{\color{nicered}\normalsize\bf}{\thesubsubsection}{1em}{}

\makeatletter							
\def\printtitle{
    {\color{bl} \flushleft \huge \@title\par}}		
\makeatother							

\title{Market structure explained by pairwise interactions}

\makeatletter							
\def\printauthor{
    {\hfill\parbox[b]{0.90\textwidth}{\flushleft \small \@author}}}				
\makeatother							

\author{%
	\textbf{\large Thomas Bury} \\[1\baselineskip]
    Service OPERA (CP194/5), Universit\'e libre de Bruxelles,\\
    Avenue F.D. Roosevelt 50, 1050 Brussels, Belgium\\
	Email:tbury@ulb.ac.be \\
	}

\usepackage{fancyhdr}
\usepackage{lastpage}	
\usepackage[runin]{abstract}		        
\abslabeldelim{:\quad}						%
\begin{document}
\printtitle

\printauthor

\begin{abstract}Financial markets are a typical example of complex systems where interactions between constituents lead to many remarkable features. Here we give empirical evidence, by making as few assumptions as possible, that the market microstructure capturing almost all of the available information in the data of stock markets does not involve higher order than pairwise interactions. We give an economic interpretation of this pairwise model. We show that it accurately recovers the empirical correlation coefficients thus the collective behaviors are quantitatively described  by models that capture the observed pairwise correlations but no higher-order interactions. Furthermore, we show that an order-disorder transition occurs as predicted by the pairwise model. Last, we make the link with the graph-theoretic description of stock markets recovering the non-random and scale-free topology, shrinking length during crashes and meaningful clustering features as expected.
\end{abstract}

\hrule
\footnotesize
\tableofcontents
\vspace{1em}
\hrule
\vspace{1em}
\normalsize
\section{Introduction}
\label{intro}

Complex systems are particularly interesting because they exhibit very sophisticated behaviors caused by, a priori, simple rules. Indeed, magnetic materials and neural networks, for instance, have some striking features such as phase transitions, memory, complicated equilibria structures and clustering. It is remarkable that these properties are caused by such simple interactions as pairwise ones.
We believe that the markets are also driven by such simple rules and that the higher-order interactions encountered in financial systems are the pairwise ones. Typical characteristics of a complex system are numerous entities and interaction rules (with a degree of non-linearity), all leading to the emergence of collective behaviors. Those behaviors in general depend more on the interactions (e.g their scaling and their order) and their effects than on the intrinsic nature of the elementary constitutive entities taken individually. The market can be viewed as such a system. The entities can be stocks or traders interacting through non-obvious rules. We note that we should interpret \emph{interaction} at the larger sense of mutual or reciprocal influence.

What one knows is that the markets exhibit features such as synchronization \cite{Dal}, structural reorganization \cite{OnnelaPRE,PeronChaos}, power laws \cite{stanley-gabaix1,stanley-gabaix2}, hierarchical and non-randomness \cite{Petra}.
What one does not know is the true market dynamics. Even if trading rules are known, microscopic equations of motion are not known. This is a fundamental difference between finance and physics (or neuroscience).

A natural approach, given the above considerations, is a statistical modeling collecting and using at best the available amount of information and allowing (in a certain sense) the emergence of critical properties. This is exactly the purpose of the maximum entropy modeling in complex systems theory.
Indeed the maximum entropy principle (MEP) allows the selection of the less restricting model on the basis of incomplete information \cite{cover}. We choose this data-based approach to avoid the use of any particular microscopic schemes (e.g. trader-agent-based rules, a priori unknown) which are difficult to assess experimentally or to avoid any analogy (even if some of such models are valuable \cite{Rosenow}). The reason is that, even if one does not know the underlying microscopic processes, the macroscopic collective behaviors can still be described by an \emph{effective} model.
One has long experience of this powerful approach in the description of phase transitions and magnetic materials \cite{ref8}. More recently, it has led to valuable results about the description of real neural networks  \cite{ref13}. Moreover, this approach also has counterparts in economics. Indeed, in addition to the  statistical meaning of the entropy, one can interpret it as a measure of the economic activity \cite{Aoki} and it is linked to the central concept of \emph{utility} of many interacting economic entities \cite{Brock,mas}.

An important outcome of such a modeling is a convenient simplified version of the real interaction structure that is still consistent with the data. In the following, we derive the model from this point of view and we study the structural properties of the resulting complex network. The critical properties will be investigated in another work.

The paper is organized as follow. In section \ref{sec:model}, we present the model, its economic interpretation and the link between the interaction matrix and the moments. In section \ref{sec:consistency}, we give evidence that the information embedded in the data is mostly explained by the pairwise but no higher-order interactions. In section \ref{sec:order}, we show an order-disorder transition through actual data. In section \ref{sec:int}, we highlight the properties of the interaction matrix and its link to the crises. Finally, in section \ref{sec:graph} we explain the link with the graph-theoretic approach and the topological evolution of the market network.

\section{The model}\label{sec:model}

\subsection{Model derivation}\label{ssec:mderiv}

The aim is to set up a statistical model describing the market state. This requires a way to infer the probability distribution in order to get the observables (here, the associated moments). The model will also allow the study of the market structure. All these quantities will be defined below.
We consider a set of $N$ market indices or $N$ stocks with binary states $s_{i}$ ($s_{i}=\pm1$ for all $i=1,\cdots,N$). A system configuration will be described by a vector $\textbf{s}=(s_{1},\cdots,s_{N})$. The binary variable will be equal to 1 if the associated index is bullish and equal to $-1$ if not. A configuration $(s_{1},\cdots,s_{N})$ is a binary version of the index returns.
One knows that this approximation is already useful in the description of neural populations \cite{ref13} and that neural networks are similar to financial networks \cite{Petra}. We may think that it will also be the case in finance; this will be justified a posteriori as the model gives consistent results.

We seek to establish a less structured model explaining only the measured index mean orientation $q_{i}=\langle s_{i}\rangle$ and instantaneous pairwise correlations $q_{kl}=\langle s_{k}s_{l}\rangle$. The brackets $\langle\cdot\rangle$ denote the average with respect to the unknown distribution $p(\textbf{s})$. As the entropy of a distribution measures the randomness or the lack of interaction among the binary variables, a way to infer such probability distribution knowing the mean orientations and the correlations is the maximum entropy principle. Jaynes showed how to derive the probability distribution using the maximum entropy principle \cite{ref12}; for supplementary information see \cite{cover}. It consists in the following constrained maximization:

\begin{eqnarray}\label{maxent}
  & \max &S(\textbf{s})= -\sum_{\{\textbf{s}\}}p(\textbf{s}) \,\log p(\textbf{s})  \\ \nonumber
   &\mathrm{s.t}&  \sum_{\{\textbf{s}\}}p(\textbf{s})=1,\quad \sum_{\{\textbf{s}\}}p(\textbf{s})s_{i}=q_{i},
   \quad \sum_{\{\textbf{s}\}}p(\textbf{s})s_{i}s_{j}=q_{ij} \nonumber
\end{eqnarray}

The resulting  two-agents distribution $p_{2}(\textbf{s})$ is the following

\begin{equation}
p_{2}(\textbf{s})=\mathcal{Z}^{-1}\exp\left(\frac{1}{2}\sum_{i, j}^{N}J_{ij}s_{i}s_{j}+\sum_{i=1}^{N}h_{i}s_{i}\right)\equiv\frac {e^{- \mathcal{H}(\textbf{s})}}{\mathcal{Z}}\label{Lagrange}
\end{equation}

where $J_{ij}$ and $h_{i}$ are Lagrange multipliers and $\mathcal{Z}$ a normalizing constant (the partition function). They can be expressed in terms of partial derivatives of the entropy as

\begin{equation}
\frac{\partial S(\textbf{s})}{\partial q_{i}} = -h_{i} \qquad
\frac{\partial S(\textbf{s})}{\partial q_{ij}} = -J_{ij}\label{multipliers}
\end{equation}

Thus preferences are conjugated to mean orientations and interaction strengths to pairwise correlations. Including higher-order correlations in constraints in (\ref{maxent}) could bring more information and thus decrease the maximum entropy. We will show below that this will not be the case.

The Gibbs distribution (\ref{Lagrange}) is similar to the one given by Brock and Durlauf in the discrete choice problem \cite{Brock} and to the one in stochastic models in macroeconomics \cite{Aoki}, and also to the Ising model used in description of magnetic materials and neural networks \cite{ref8,ref13}. It is also a special case of Markov random fields \cite{ref15}. It is to be noted that the Gibbs distribution and Shannon entropy naturally arise from the stochastic modeling in economics; this is discussed in \cite{Aoki}.

We obtain the parameters $\{J_{ij},h_{i}\}$  by performing explicitly the maximization (\ref{maxent}) so that the theoretical moments $\langle s_{i}\rangle$ and $\langle s_{i}s_{j}\rangle$ match the measured ones $q_{i}$ and $q_{ij}$. We note that this requires the computation of $2^N$ terms. If this number is large, the computation will take a while and we can benefit from one of the methods described in \cite{ref4}.

Last, we show how the cumulants are obtained from this model and their relation to interaction strengths. As the statistical model (\ref{Lagrange}) is expressed as a Gibbs distribution, we have the relations

\begin{equation}
\langle s_{i_{1}}\ldots s_{i_{N}}\rangle_{\mathrm{c}}=\partial^{N}\ln \mathcal{Z}/\partial h_{i_{1}}\ldots \partial h_{i_{N}}
\end{equation}

where $\langle\cdot\rangle_{\mathrm{c}}$ is the cumulant average \cite{Kubo}; it gives the relation between $\mathbf{J}$ and the correlation functions. If the partition function $\mathcal{Z}$ cannot be explicitly computed, we can use the Plefka series \cite{Plef} or a variational cumulant expansion \cite{Barb}.

Hereafter, we will show that the covariances are consistently deduced from this statistical model and thus that they are a function of the interaction strengths.

\subsection{Interpretation}\label{ssec:interpr}

We interpret the objective function $\mathcal{H}(\textbf{s})$ defined by the MEP  in the distribution (\ref{Lagrange}) as follows.
The pairwise interactions between economic agents are modeled by interaction strengths $J_{ij}$ which describe how $i$ and $j$ influence each other. Here by \emph{interaction}, we mean a measure of mutual influence or a measure of share comovement. In this framework, our intention is not to give a description of these interactions but to study their effects. Actually, the causes underlying the interaction process seem to be unnecessary in the description of emergent macroscopic behaviors. Indeed the complicated interactions between magnetic moments or between neurons are efficiently simplified in their maximum entropy description but one still recovers the main macroscopic features observed in these systems. In this description, the crucial features are the scaling (dependence on or independence of the system size) of interaction strengths and the order of interactions. The matrix $\textbf{J}$ is set to be symmetric in this first approach. There is disagreement or conflict between entities when the weighted product of their orientations $J_{ij}s_{i}s_{j}$ is negative. If two shares are supposed to move together ($J_{ij}>0$), a conflicting situation is the one where they do not have the same orientation (bearish or bullish).

We include the idiosyncratic preferences of the economic agents, here the willingness to be bullish or not. These Lagrange multipliers $h_{i}$ can also be interpreted as the external influences on entities $i$ induced by the macroeconomic background. By example a company can prosper and make benefits during a crisis period and the associated stock can still fall simultaneously because the investors are negatively influenced by the economic background. It results that the stock will have a propensity to fall. We denote the external influence by $h_{i}$. If $i$'s orientation satisfies its preference, $h_{i}s_{i}$ is positive.
The total conflict of the system is thus given by

\begin{equation}\label{Hfunc}
\mathcal{H}(\textbf{s})=-\frac{1}{2}\sum_{i=1}^{N}\sum_{j=1}^{N}J_{ij}\, s_{i}s_{j}-\sum_{i=1}^{N}h_{i}s_{i}
\end{equation}

So, we interpret $\mathcal{H}(\textbf{s})$ as the opposite of the so-called utility function $\mathcal{U}(\textbf{s})=-\mathcal{H}(\textbf{s})$ with pairwise interacting and idiosyncratic parts. Consequently the interaction strengths can be viewed as the incentive complementarities \cite{Brock,mas}. Indeed we have $\partial^{2} \mathcal{U}/\partial s_{i}\partial s_{j}=J_{ij}$.The larger $J_{ij}s_{i}s_{j}$ is, the stronger the strategic interaction between $i$ and $j$ is.

We emphasize that this Ising like model is forced upon us as the statistically consistent model with the measured orientations and correlations. It is not an analogy based on specific hypotheses about the market dynamics.

\section{Consistency of the pairwise modeling}\label{sec:consistency}
\begin{figure}[!ht]
\begin{center}
\resizebox{0.85\textwidth}{!}{%
\includegraphics[height=3cm]{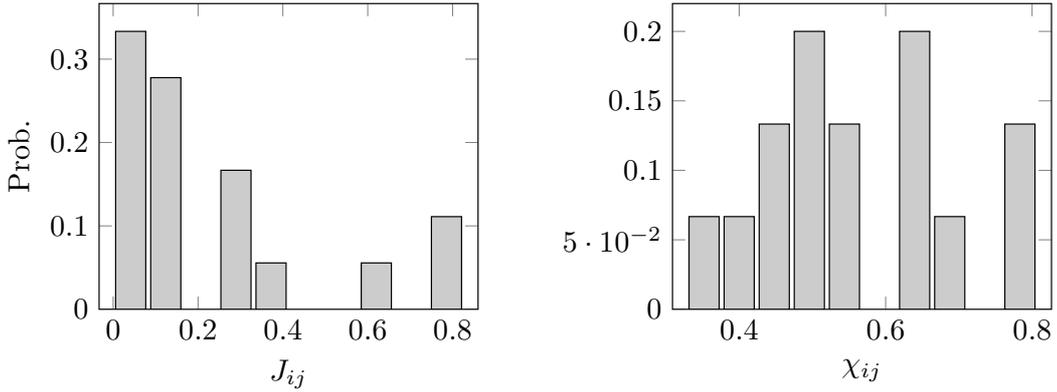}
 }
\end{center}
\caption{\label{fig:prob} Left: maximum entropy distribution of the interaction strengths $\tilde{\textbf{J}}^{\mathrm{\scriptstyle{ME}}}$ and right: empirical distribution of the pairwise correlations obtained from the collected data.}
\end{figure}

One of the most exciting features of the model is the emergence of collective behaviors even if the interactions are weak. If the model is able to explain the recorded data, the system is therefore dominated by pairwise correlations.
The aim is to provide quantitative empirical evidence that the pairwise modeling is a consistent paradigm to explain the financial data and exhibited behaviors in the market.
In the following, we apply the pairwise model to a set of six major market indices (AEX, Bel-20, CAC 40, Xetra Dax, Eurostoxx 50, FTSE 100). We selected only European indices because some financial issues are specific to Europe and we consider indices because they are the driving force of the respective stock markets \cite{Shap}, they will reflect the main properties of the subjacent stock set. We will say that they are \emph{up} or \emph{bullish} if the closing price is higher than the opening price and they are \emph{down} or \emph{bearish} if not. These will constitue our binary states. We observe 2253 configurations from 6/06/2002 to 14/06/2011 \cite{ref3}. We take a nine year long time series including two large crises. The daily sampling is enough since we want to study large crises, and the two principal peaks of the Fourier transform are centered on frequencies $f_{1}=6\times10^{-4}\,\mathrm{d}^{-1}$ and  $f_{2}=1.2\times10^{-3}\,\mathrm{d}^{-1}$; the unit \emph{day} stands for trading day. The first frequency $f_{1}$ is the crisis occurrence frequency in our time window, the corresponding period is $T_{1}=1.7\times10^{3}\,\mathrm{d}$ . Later, we will also analyze the stocks composing the Dow Jones and the S$\&$P100 indices, and another set of 116 stocks.
First of all, we give the magnitude order of the interaction strengths and of the empirical pairwise correlations in Fig-\ref{fig:prob}.


The $J_{ij}$ are all positive; we can therefore use net mean orientation (net magnetization) as an order parameter. The mean value of $h_{i}$ is about $0.0113$.

As mentioned above, higher-order interactions can be involved in the interaction structure. In order to show that pairwise correlations are prevailing, we compute the Kullback-Leibler (KL) divergence, $D_{\scriptstyle{\mathbf{KL}}}(P_{2}\|P_{\scriptstyle{\mathbf{data}}})$ between the two-agents maximum entropy (ME) distribution $P_{2}$ and the empirical one $P_{\scriptstyle{\mathbf{data}}}$. The KL divergence is equal to $2.27\times 10^{-2}$ for the ME distribution inferred from 2253 observations. It must be compared to $D_{\scriptstyle{\mathbf{KL}}}(P_{1}\|P_{\scriptstyle{\mathbf{data}}})=1.4801$ for the independent agents model $P_{1}$. The closer to zero this quantity is, the closer $P_{2}$ to $P_{\scriptstyle{\mathbf{data}}}$ is. Specifically, a consistent way to test if the pairwise correlation model satisfactorily explains data statistics is to evaluate the ratio between $S(P_{1})-S(P_{2})$ and the Kullback-Leibler discrepancy $I_{N}\equiv D_{\mathrm{KL}}\left(P_{N}||P_{1}\right)$, where $S(P_{2})$ is the entropy of the pairwise model. If this ratio is close to 1, the pairwise correlations explain most of available information. Indeed the multi-information $I_{N}=S(P_{1})-S(P_{N})$ measures the total amount of correlations in the system \cite{ref16}.
In this application, we obtain $I_{2}/I_{N}=98.5\%$. The pairwise correlations model is effective since it explains almost all the available information; only $1.5\%$ of information is due to higher-order interactions.

\begin{figure}[!ht]
\begin{center}
\includegraphics[width=0.5\textwidth]{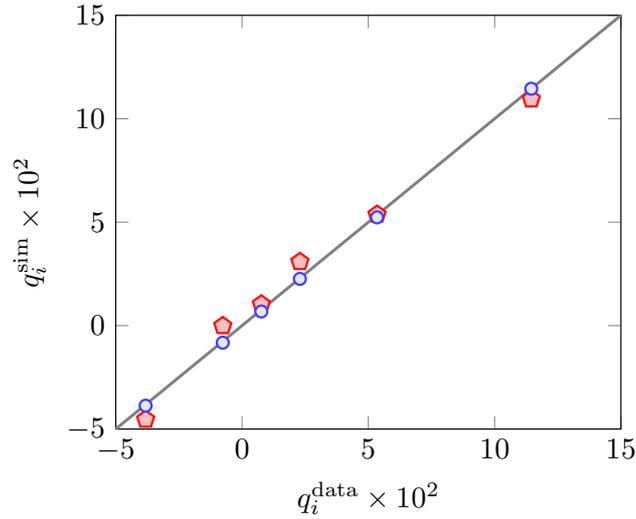}
\end{center}
\caption{Comparison of simulated orientations and the actual ones. The straight line shows equality. The circles stand for simulations with exact Lagrange parameters and pentagons stand for approximated mean-field Lagrange parameters.}
\label{fig:comp}
\end{figure}

As a further test of the pairwise model consistency, we show below that this statistical model is able to recover the observed empirical moments. We compare the average index orientations $q_{i}=T^{-1}\sum_{t=1}^{T}s_{i,t}$ obtained by simulation to the real ones. We simulated the process by doing $1\times 10^{5}$ equilibration Monte Carlo time steps (MCS) and we take the average on the next $2\times 10^{7}$ MCS in order to reduce the variance of the estimator. The flipping attempts are simulated by the Glauber dynamics. Namely, we take an entity $i$ chosen randomly and the attempt to flip the associated binary variable $s_i$ is performed with a rate depending on the exponential weight, the other orientations remaining fixed \cite{ref2}. We take the time average for each index from the data and we compare it to the value obtained with the simulation; they are illustrated in Fig-\ref{fig:comp}.


\begin{figure}[h!]
\begin{center}
\includegraphics[width=.75\textwidth]{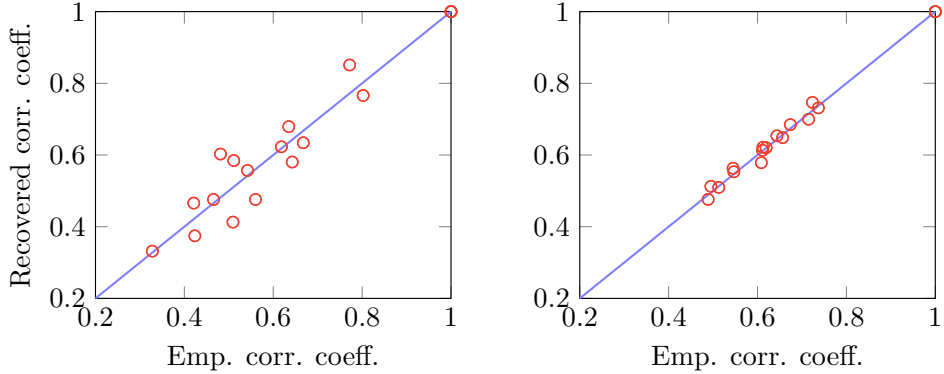}
\end{center}
\caption{(Color online) Recovered correlation coefficients from MCMC versus empirical ones. The straight line shows equality. The result based on 130 observations (left) and the result based on 2258 observations (right).}
\label{fig:corr}
\end{figure}

The root mean squared error (RMSE) is equal to $7\times 10^{-4}$, which represents $1.5\%$ of the root mean squared (RMS) value of the six arithmetic means (equal to $4.90\times 10^{-2}$). We recover quantitatively the average orientation of the six indices on the observation period. Moreover, since we obtained the probability distribution, we can compare the correlation coefficients resulting from the sampling of the proposed probability distribution to the empirical ones. We sample the probability law $p_{2}(\textbf{s},\mathbf{J}^{\mathrm{\scriptstyle{ME}}},\mathbf{h}^{\mathrm{\scriptstyle{ME}}})$ by a Monte Carlo Markov chain (MCMC). We take $1.2\times 10^{6}$ equilibration steps and $1.2\times 10^{4}$ independent sampling steps between each sample. Fig-\ref{fig:corr} illustrates the recovered correlation coefficients with the maximum entropy estimation versus the empirical ones. The results for only 130 observations (chosen arbitrarily corresponding to half a year) are conclusive. Indeed the RMSE represents $8.3\%$ of the RMS value and the correlation coefficient of the empirical and simulated values is equal to $0.963$. Including more observations (2258 trading days) allows us to reduce the dispersion in the results (correlation coefficient of the empirical and simulated values equal to $0.997$; the RMSE represents $1.8\%$ of the RMS value). We note that it is effective even with few data.

\begin{figure}[ht!]
\begin{center}
\resizebox{1\textwidth}{!}{%
  \includegraphics{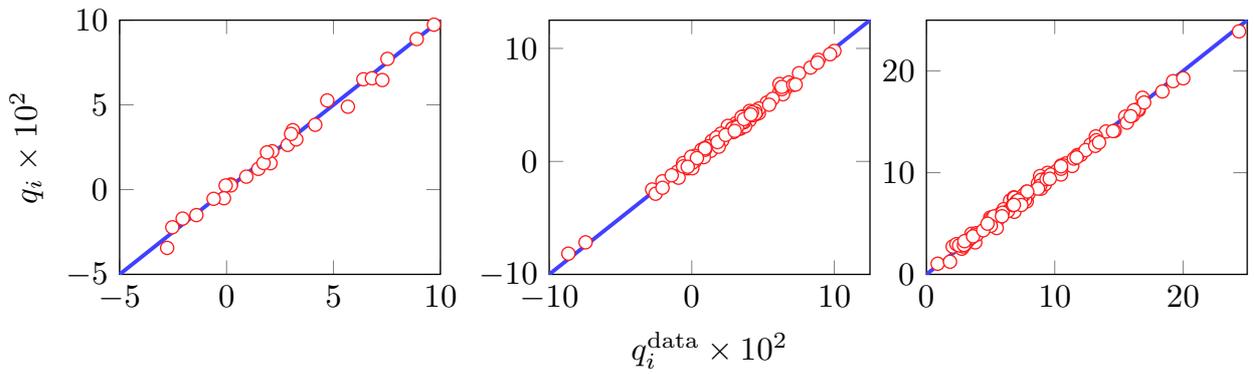}
}
\end{center}
\caption{(Color online) Comparison of simulated orientations and the actual ones. From left to right: DJ, S$\&$P100 and Onnela's set. The straight line shows equality.}
\label{fig:choicesAl}
\end{figure}

\begin{figure}[h!]
\begin{center}
\resizebox{\textwidth}{!}{%
  \includegraphics{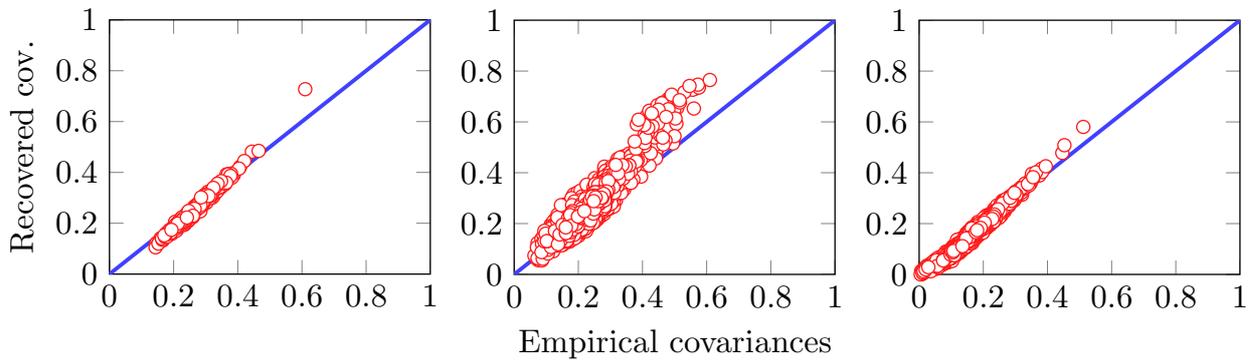}
}
\end{center}
\caption{(Color online) Recovered covariances versus empirical ones. From left to right: DJ, S$\&$P100 and Onnela's set. The straight line shows equality.}
\label{fig:covAll}
\end{figure}

We perform the same work for the Dow Jones and the S$\&$P100 indices (2500 configurations observed from 10/10/2001 to 02/08/2011). We also consider 116 stocks from the New York Stock Exchange available on the Onnela's website (http://jponnela.com/) extending from the beginning of 1982 to the end of 2000 (4800 trading days). For these larger stock sets, the exact entropy maximization (\ref{maxent}) is not computationally tractable. There are several approximate inversion methods to estimate the parameters. The mean field methods (na\"{\i}ve, TAP and Tanaka's inversion see \cite{ref4,Sc,refTanaka}) are the faster ones and they are accurate if the interaction strengths are weak (the \emph{weakness} will be investigated in a further work). These methods will be used in the investigation of the structure evolution due to their reasonable accuracy and quickness. Two other valuable inference methods are minimum probability flow (MPF) \cite{Sohl} and regularized pseudo-likelihood maximization (rPLM) \cite{Aurell}. In our application the rPLM method performs best.
The results for the first and second recovered moments ($2\times10^{6}$ equilibration MCS, values estimated on $2\times10^{7}$ samples recorded each $N$ MCS) are illustrated in Fig-\ref{fig:choicesAl} and Fig-\ref{fig:covAll}.

The correlation coefficient between the recovered and empirical values is respectively $0.998$, $0.996$ and $0.997$ for the net orientations illustrated in Fig-\ref{fig:choicesAl} and $0.989$, $0.964$, $0.997$ for the covariances illustrated in Fig-\ref{fig:covAll} which shows the strong linear statistical relation between the empirical and the recovered values.
The relative deviation between the RMSE and the RMS values is respectively $2\%$, $7\%$ and $6\%$ for the net orientations and $9\%$, $17\%$, $8\%$ for the covariances.

We have seen that, in addition of the multi-information criterion, the net orientations and the covariances are recovered from this model even with few data. We conclude that the proposed pairwise interaction structure is a trustful one; this means that interactions are believed to be pairwise and symmetric ones and that they cause correlations.

\section{Order-disorder transition}\label{sec:order}
\begin{figure}[h!]
\begin{center}
 \includegraphics[width=\textwidth]{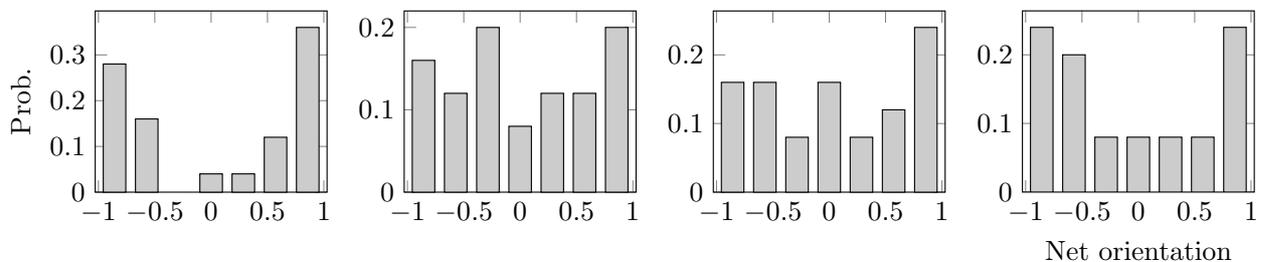}
\end{center}
\caption{Empirical probability distribution of the net orientation on four successive periods, each of 25 trading days. Time goes from left to right. The last time window corresponds to the irregularity induced by the Fukushima nuclear accident.}
\label{fig:EmpTrans}
\end{figure}

As the previous pairwise model describes market indices quantitatively, we expect to observe an order-disorder transition in this system; we give below some empirical evidence that these transitions actually appear.
As the interaction strengths are all positive, the system is ordered if the net orientation distribution has two modes near the extreme values $-1$ and $1$ and disordered if the distribution has a unique mode. Indeed in an ordered situation, each index tends to have the same orientation as the others. Furthermore, in the absence of external influences, both extreme values are equivalent (as a consequence of the symmetry under sign exchange), and the distribution is thus bimodal. One of the extreme values can be favored following the values taken by the external influences $h_{i}$. It will be a first clue that the system is reorganized if the distribution changes in such a way (having two modes and then a unique one, and reciprocally).
We compute the system net orientation $q(\tau,\Delta t)=(\Delta t\,N)^{-1}\sum_{i}\sum_{t=\tau}^{\tau+\Delta t}s_{i,t}$ on successive periods $\Delta t$ of 25 trading days (without overlapping), and we show that the net orientation probability distribution can be bimodal or not on successive time windows. The resulting empirical distributions for observations from 5 November 2010 to 30 March 2011 are illustrated in Fig-\ref{fig:EmpTrans}.

In Fig-\ref{fig:EmpTrans} we see that the empirical probability distribution has initially two modes at extreme orientation values then has no clear mode, and finally again has two modes. During this period, initially the indices move in an organized way then in a disorganized fashion, and finally the Fukushima nuclear accident caused a large global market fall followed by a large recovery. During this event, the indices were in comovement. So the system is initially ordered then disordered for two periods and then again ordered.

Another way to characterize financial irregularities is to study the entropy $S(\textbf{s})$ on a sliding window (here, 300 trading days shifted by 1 day). We compute the mean-field approximation of the entropy on those time windows (much faster than the exact computation). The mean-field entropy \cite{binder} is

\begin{equation}\label{MFentr}
  S_{\mathbf{MF}}(\textbf{s})= -\sum_{i=1}^{N}\frac{1+q_{i}}{2}\ln(\frac{1+q_{i}}{2})+
  \frac{1-q_{i}}{2}\ln(\frac{1-q_{i}}{2})
\end{equation}

\begin{figure}[h!]
\begin{center}
\includegraphics[width=0.75\textwidth]{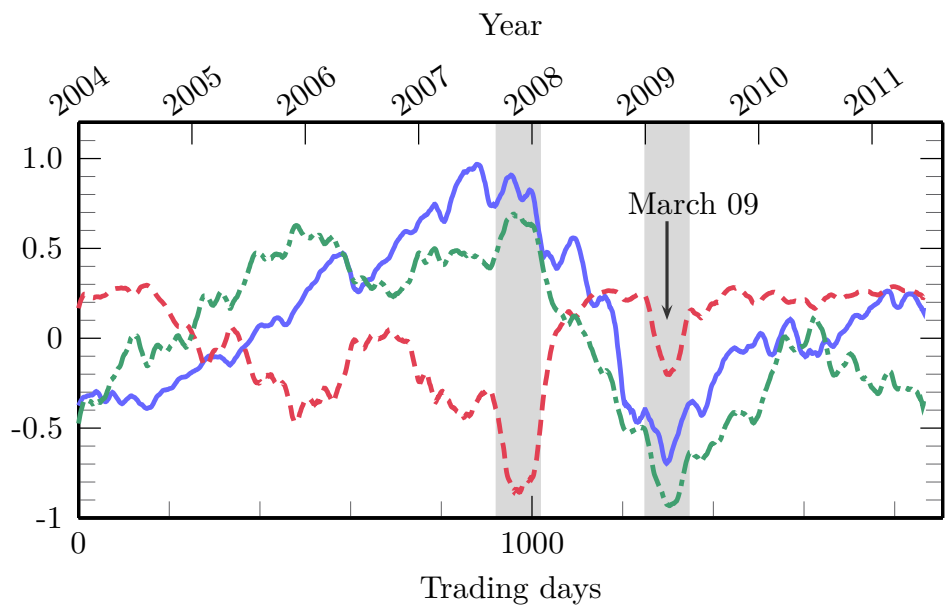}
\end{center}
\caption{The normalized sum of indices (full line), the normalized net orientation (dashed-dotted line) and the normalized mean-field entropy (dashed line). The curves have been smoothed. The last major crisis is pointed out by an arrow. The shaded portions show orientation extrema and entropy minima. }
\label{fig:EurEntrSmooth}
\end{figure}

The entropy is maximal when the average orientations, computed on the corresponding time window, are  equal to zero and is minimal when the indices have the same orientation. During a disordered period, the entropy should be large and during a synchronized (ordered) period the entropy should be low. We should thus observe entropy minima simultaneously to orientation extrema (bubbles or crashes). We check in the results illustrated in Fig-\ref{fig:EurEntrSmooth} that orientation extrema and entropy minima are related to the periods of synchronization described in \cite{Dal}.

We observe large falls of the entropy when the net orientation is much larger than its mean (the mean is set to zero in Fig-\ref{fig:EurEntrSmooth}). The shaded portions show the orientation extrema and entropy minima on this time window. They correspond (chronologically) to the end of the growth period and the end of the collapse. Furthermore the correlation coefficient of the net-orientation and the financial time series is equal to $0.82$ showing a high degree of linear statistical dependency. We conclude that the entropy minima are thus related to financial irregularities (large upward or downward movements).

This is an empirical evidence that order-disorder transitions occur in markets. This interpretation is supported by the recent results obtained in \cite{Dal}, where the authors showed that market irregularities present a high degree of synchronization, meaning an ordered state.
The economic consequence is that the whole market is correlated when such transitions occur. It also means the absence of a characteristic scale for the fluctuations and the emergence of power-laws.

In appendix, we illustrate in Fig-\ref{fig:EurEntr} a larger version of Fig-\ref{fig:EurEntrSmooth}.

\section{Dynamics of interactions}\label{sec:int}

Linked to the above, such a transition occurs if the stochasticity changes or the interaction strengths change. A possible interpretation of time-varying interaction strengths is that some learning or adaptive process takes place through time. This means that the market adjusts the interactions between its entities in some adaptive processes so the $\{J_{ij},h_{i}\}$ are time dependent. The reason is that the background, namely worldwide economic conditions, changes through time and goes through economic fluctuations with contractions (recessions) and expansions (growths). As the correlations are explained by the pairwise interactions, it also means that the correlations to be do not necessarily match past correlations.

\begin{figure}[h!]
\begin{center}
  \includegraphics[width=0.75\textwidth]{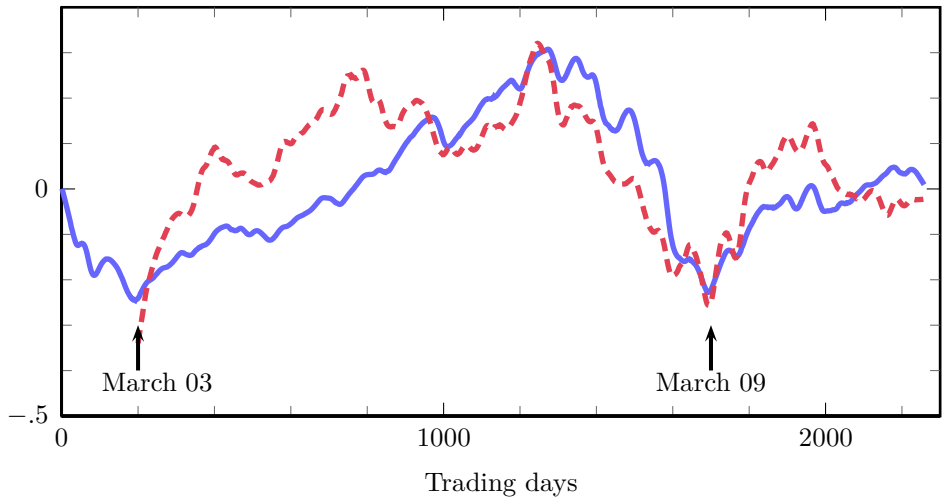}
\end{center}
\caption{The aggregate preference (dashed line) and the normalized sum of indices (full line); both curves have been smoothed. The last two major crises are pointed out by arrows. }
\label{fig:EurPref}
\end{figure}

Following this interpretation, we expect that the temporal behaviors of the interaction strengths and external influences are related to market evolution. This is indeed true, as we will see below. First of all, we study the preference evolution of the six previous indices (reflecting the current state of the European economy) and its link to the crises.
We use a sliding temporal window of width $T=200$ trading days shifted by a constant amount of  $\Delta t=2$ trading days. We show that the aggregate preference $h=\sum_{i}h_{i}$ is negative during a crisis (or during a significative contraction) as illustrated in Fig-\ref{fig:EurPref}.

The first negative incursion corresponds to the 2002-2003 crisis and the second one to the 2008-2009 crisis \cite{ref3}.
As expected the external influences are decreasing when the market is subject to a crisis.

\begin{figure}[h!]
\begin{center}
\includegraphics[width=0.75\textwidth]{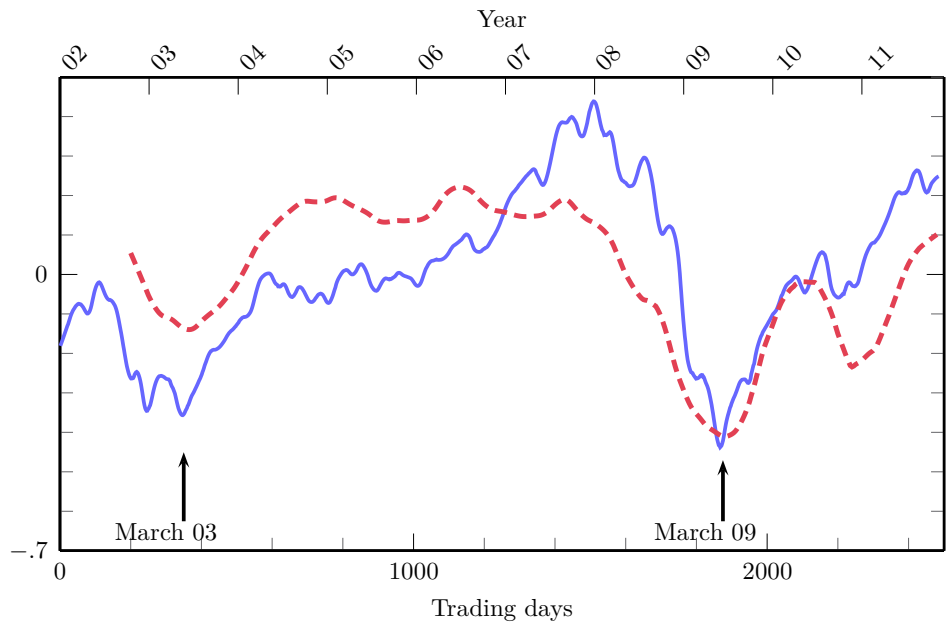}
\end{center}
\caption{The normalized Dow Jones index is plotted as a full line; the trace minus its temporal average is the dashed line. We used a sliding temporal window of width equal to 200 trading days shifted each time by 5 trading days.}
\label{fig:DJspectre}
\end{figure}

More interestingly, we will study the spectrum of the interaction matrix. Indeed the spectrum evolution will be related to the market evolution. The spectrum of the interaction matrix of a stock set has an interesting feature; we will show it for the Dow Jones index. We collected data for the Dow Jones index from the 10 October 2001 to 1 August 2011 \cite{ref3}, and we extract the interaction strengths using the third-order approximation described in \cite{refTanaka}.
The trace of the interaction matrix, the sum of its eigenvalues, has the following interesting property. It decreases during a crisis; specifically, the trace minus its temporal average becomes negative if there is a substantial fall of the index, this feature is illustrated in Fig-\ref{fig:DJspectre}.

The trace of the exact interaction matrix should be zero (without self-interactions) but, with the Tanaka's diagonal trick, the diagonal entries are related to the second-order term and to a part of the third-order of the Plefka series \cite{refTanaka,Plef}. The second-order term of the Plefka series is negative, the sign of the third-order term depends on the product of the interaction strengths. The temporal variation of the trace reflects the temporal variation of these second and third order terms. These terms are particularly important near a transition. This explains why the trace of the interaction matrix is smaller than its mean value during a crisis. Indeed during a crisis all the stocks act in similar way: they fall down. They thus have similar mean orientation (down) and the resulting system state is an ordered one. Before the crisis, during a common market growth or steady state, the price of some stocks rises (on average) and some others fall leading to a dispersion of the mean orientations. This is indirect evidence of a transition from one regime to another and of coordination. It is consistent with the results obtained above and in \cite{Dal,Jr}. In appendix, we illustrate in Fig-\ref{fig:DJspecHuge} a larger version of the Fig-\ref{fig:DJspectre}.

\section{Link to the graph-theoretic approach}\label{sec:graph}

\begin{figure}[h!]
\begin{center}
\includegraphics[width=0.75\textwidth]{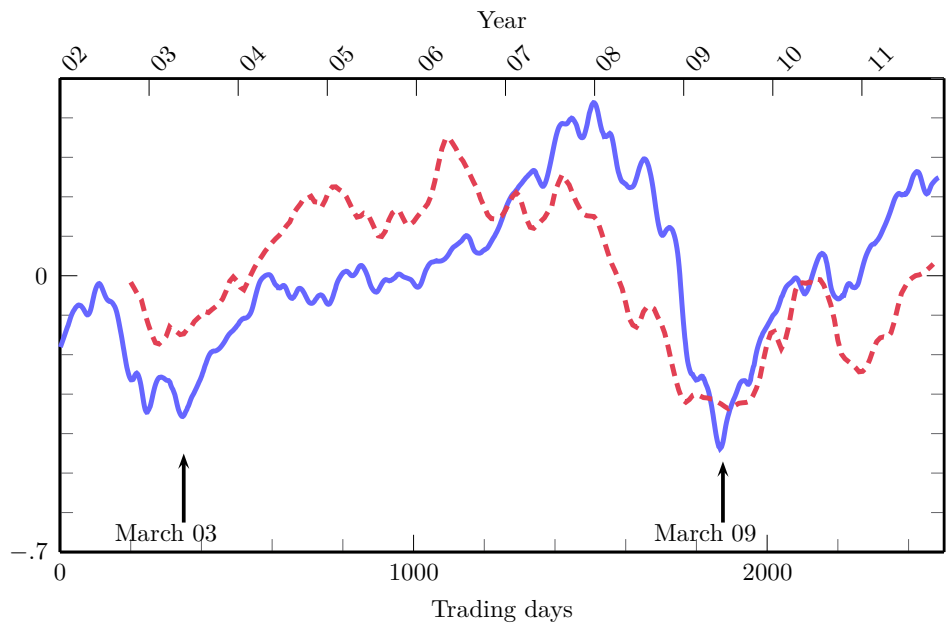}
\end{center}
\caption{The normalized Dow Jones index is plotted in full line, the relative difference to the time average of the length $l(t)=\left[L(t)-\langle L\rangle\right]/\langle L\rangle$ is the dashed line (where the brackets denote the temporal average). We use a sliding window of 100 trading days shifted by 10 trading days each time.}
\label{fig:DJmst}
\end{figure}

Hereafter, we make the link with the previous spectrum feature and the observation that the length of the minimum spanning tree (MST) based on the Sornette-Mantegna distance decreases during a crash \cite{Mant,Onnela}, meaning that stocks are highly correlated during these events (as they should be in an order-disorder transition). We will see that we recover this feature with the pairwise model with a distance based on interaction strengths in place of correlation coefficients. Indeed the interaction matrix can be thought of as the weight matrix of an undirected complete graph. Using a modified version of the method proposed in \cite{Ding} and computing the minimum spanning tree length $L(t)$ (the sum of the edges weights of the MST), we also observe that this length decreases during a crash, as expected; the results for the Dow Jones index are illustrated in Fig-\ref{fig:DJmst}.

\begin{figure}[h!]
\begin{center}
\includegraphics[width=\textwidth]{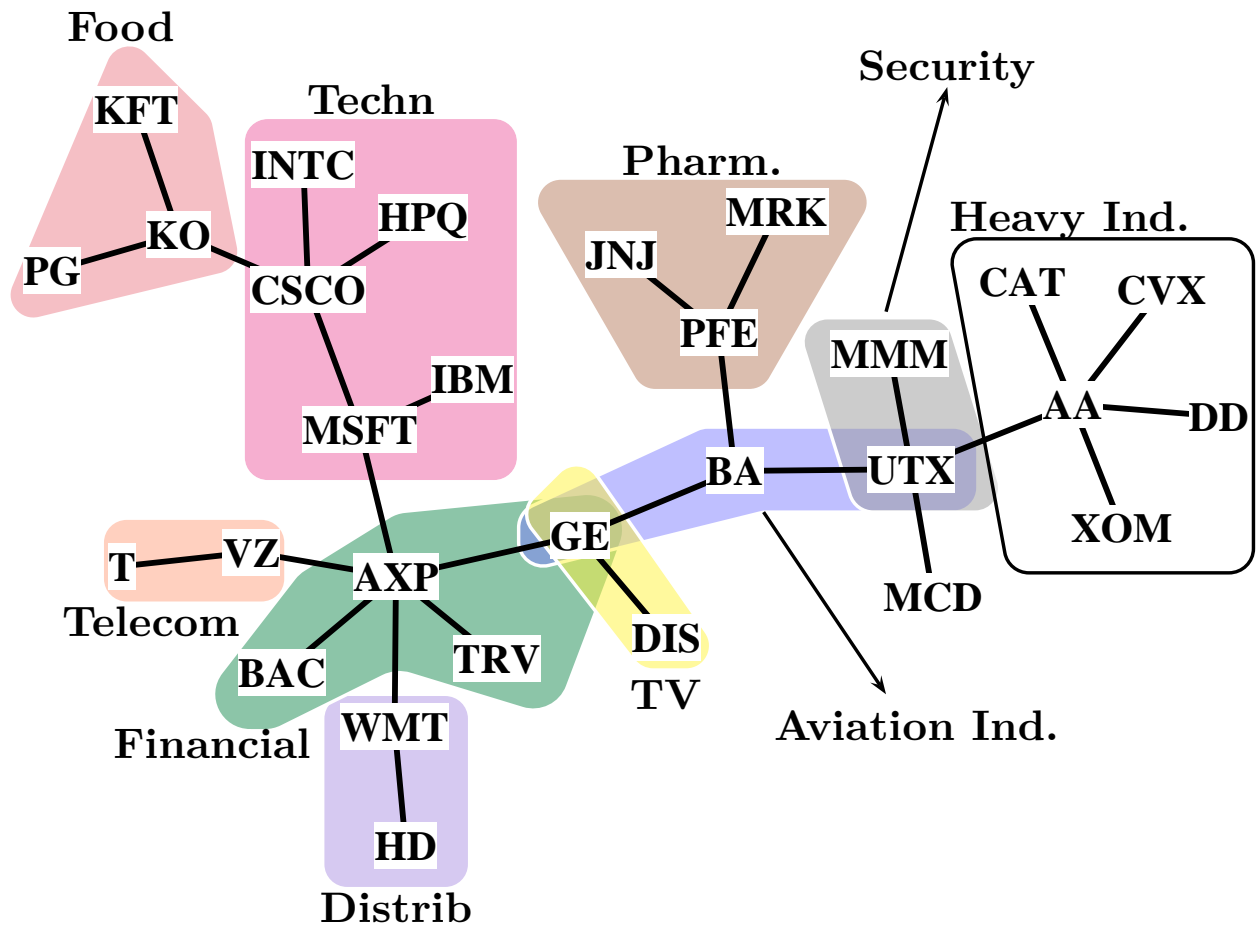}
\end{center}
\caption{(Color online) The minimum spanning tree based on the interaction matrix $\mathbf{J}$ is estimated on 2500 trading days. The companies are denoted by their ticks; they can be found on any financial website (\emph{Google finance} for instance).  }
\label{fig:DJclusters}
\end{figure}

Moreover, it also allows cluster identification. Indeed, it is known that the asset tree based on the Sornette-Mantegna distance allows regrouping some stocks in clusters following their economic sectors \cite{Onnela}. As the correlations are caused by the interactions, it is not surprising that the MST of the network defined by the interaction matrix also allows cluster identification. This approach has the advantage of not being limited to linear or monotonic statistical dependencies. The clustering feature is illustrated in Fig-\ref{fig:DJclusters}.

We note that General Electric (GE) is not the most connected node but it is a cental one in the sense that it appears in three different clusters, as such it is still considered as the \emph{root} of the MST and defines the generational direction. This approach provides a different classification than the one given in \cite{Onnela} or given by Forbes for instance. Indeed, Forbes classification is given by sector then by industry. Disney and Walmart are classified in the same sector, \emph{services}; this category is too vague to be an useful tag. Similarly, General Electric is tagged by Forbes as \emph{industrial goods} and then as \emph{diversified machinery} but this company also provides financial services, aircraft engines, TV channel broadcasting, etc. It is then clear that this company should be classified with more than one tag, as does the proposed method. In this point of view, the internal structure of each company seems to be the crucial information to identify stock clusters.

\begin{figure}[h!]
\begin{center}
\includegraphics[width=0.25\textwidth]{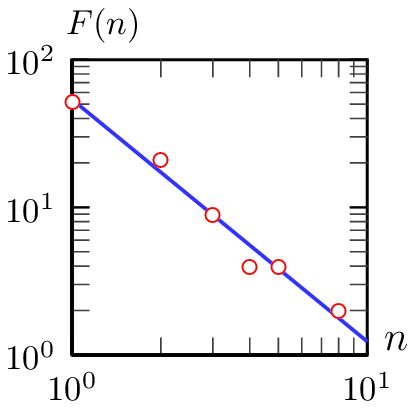}
\includegraphics[width=0.25\textwidth]{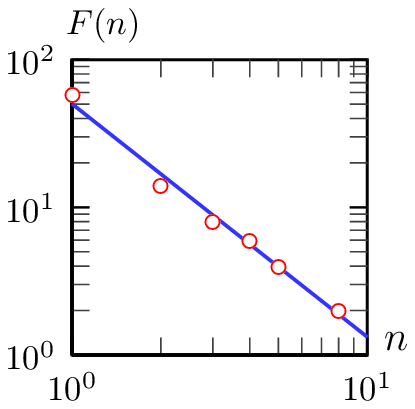}
\end{center}
\caption{The degree distributions during a growth period (left) and during a crash (right). The solid line is a power-law fit; the coefficients of determination are respectively $0.98$ and $0.93$.  }
\label{fig:SPdegree}
\end{figure}

We can also study the topological structure of the remaining asset tree during a crash and a growth period. We will see that, as expected, the degree distribution follows a power law. We consider the stocks of the S\&P100 index on two intervals, from 1/10/2007 to 01/02/2009 (360 trading day crisis period) and from 1/02/2005 to 1/07/2007 (600 trading day growth period). The occurring frequencies of the vertex degrees are illustrated in Fig-\ref{fig:SPdegree}.

This reveals that the degree distribution is a power law, $f(n)\sim n^{-\alpha}$, and the value of the exponent is similar for the both periods. For the growth period, we obtain $\hat{\alpha}=1.64\pm 0.17$ and during a crash $\hat{\alpha}=1.58\pm 0.12$. They can be included in the confidence interval of each other, so they are very similar.
The maximum degree is $n=8$ in the both periods. They are 58 vertices of degree $n=1$ during the crash. This value is slightly larger (about $10\%$) than the one corresponding to the growth period, 52 vertices of degree $n=1$. This explains the difference between both exponents. The asset tree topology is thus slightly different during a crash. The main change is the variation of the interaction strengths (the graph weights) rather than the variation of the vertex degrees.  In both regimes, the asset trees are thus scale-free networks. This implies that the edges are not drawn at random and the asset trees exhibit small-worldness, as observed with another method in \cite{Petra}. Furthermore, the low value of this exponent implies that hubs (high-degree vertices)  represent a significant part of the total number of vertices. The market is thus sensitive to the failure of a hub (a highly connected company) whereas the failure of a leaf (terminal node) will only slightly affect the market. By example the hypothetic failure of the American Express Company (AXP) would leave a fragmented market whereas the bankruptcy of Kraft Food Inc. (KFT) would not change the topology of the asset tree significantly; see Fig-\ref{fig:DJclusters}. This could help in selecting the companies one has to save from an eventual bankruptcy in order to minimize the impact of such an event. This could also help to select which companies one has to monitor to prevent a hypothetical dramatic system failure.

\section{Conclusion}
We have seen that, without making assumptions on the market dynamics, the maximum entropy principle provides a rigorous pairwise model which is able to describe the data and the observed collective behaviors quantitatively. We showed that including higher-order interactions does not explain more than using the pairwise model, and thus that the collective phenomena emerge from simple pairwise interactions. To confirm this result, we showed that this statistical model is able to recover the empirical moments computed from the data, especially the mean orientations and the correlations.
The success of the pairwise model implies that markets exhibit some properties observed in magnetic materials and in neural networks. Indeed, we showed that an order-disorder transition occurs in such a system, as described by a pairwise model equivalent to the Ising model. Furthermore, we showed that the interaction strengths are time dependent meaning that an adaptive process occurs and that they are the starting point of the graph-theoretic approach of the market.
In this view the system is more than the sum of its parts, is ruled by its entities pairs, exhibits collective behaviors and is quantitatively described by a pairwise model. It is surprising that such sophisticated collective behaviors, emergent structures and underlying complex trading rules are captured by a simple (a priori) scheme of interdependence involving only pairwise but no higher-order interactions.

\section*{Acknowledgments}
I would like to thank B. De Rock and P. Emplit for their helpful comments and discussions. This work was undertaken with financial support from the Solvay Brussels School of Economics and Management.

\appendix

\begin{figure}[ht!]
\begin{center}
\resizebox{\textwidth}{!}{%
  \includegraphics{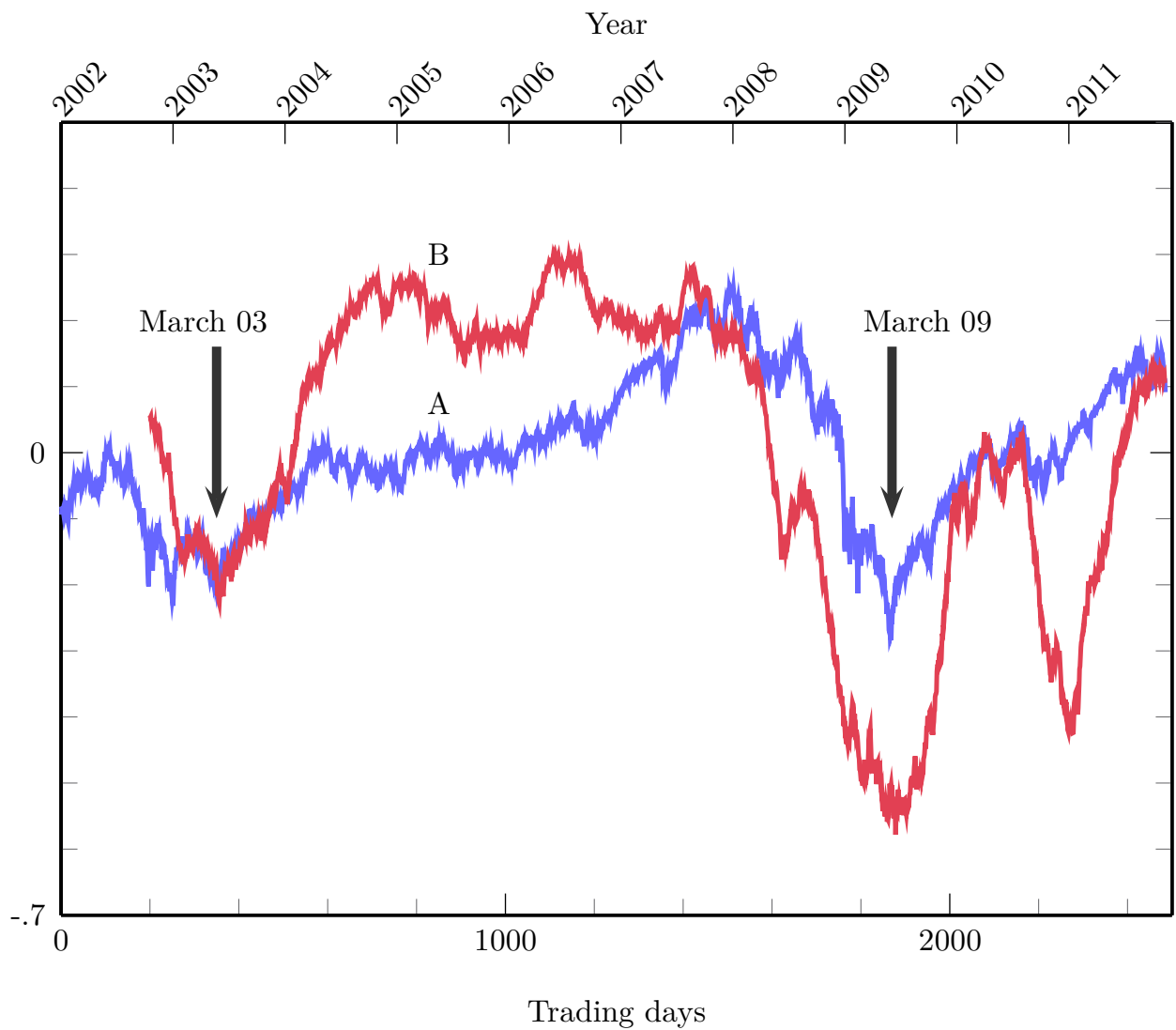}
}
\end{center}
\caption{The normalized Dow Jones index is plotted (curve A); the trace minus its temporal average is the dashed line (curve B). We used a sliding temporal window of width equal to 200 trading days translated each time by 1 trading day.}
\label{fig:DJspecHuge}
\end{figure}

\begin{figure}[ht!]
\begin{center}
\resizebox{\textwidth}{!}{%
  \includegraphics{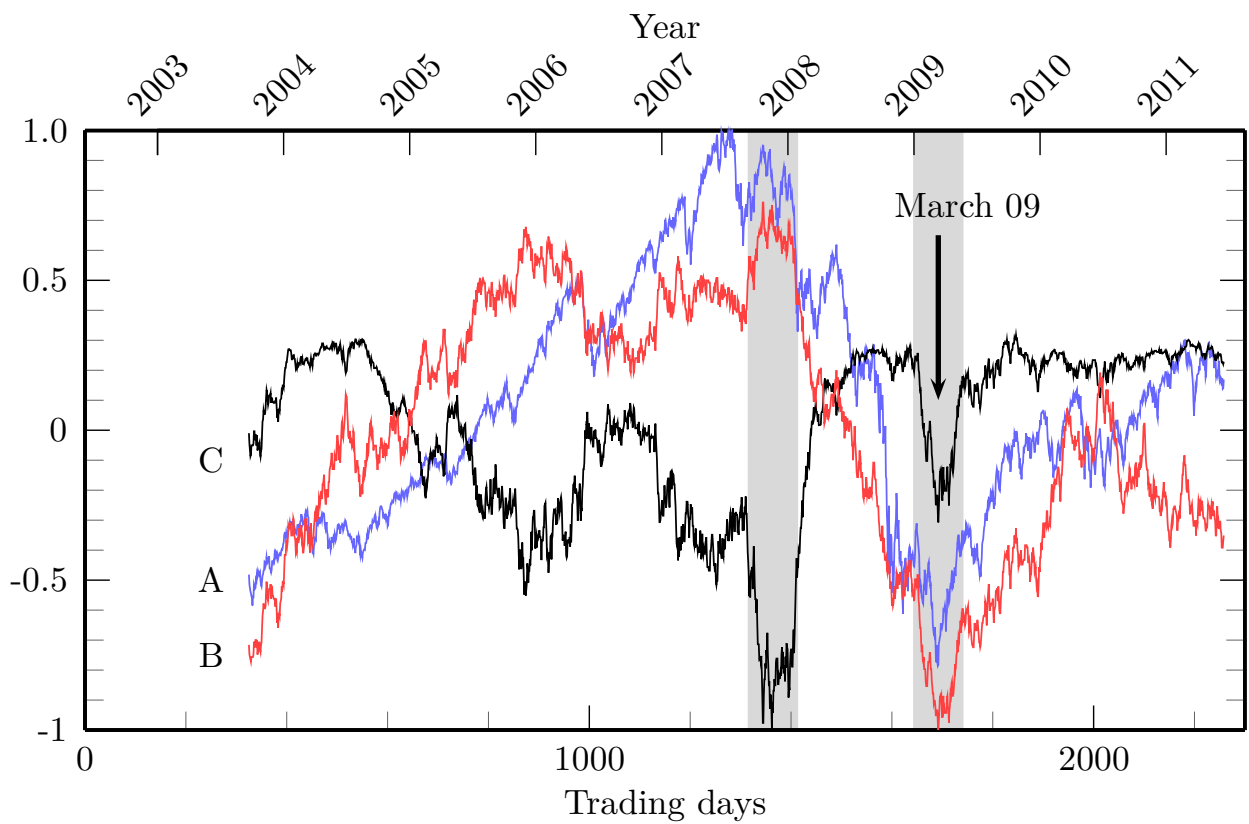}
}
\end{center}
\caption{The normalized sum of indices (curve A), the normalized net orientation (curve B) and the normalized mean-field entropy (curve C). The last major crisis is pointed out by an arrow. The shaded portions show orientation extrema and entropy minima.}
\label{fig:EurEntr}
\end{figure}

%

\end{document}